\documentclass{optica-article}

\journal{opticajournal} 

\articletype{Research Article}

\usepackage{lineno}
\usepackage{bm}
\usepackage{titlesec}
\usepackage{siunitx}

\begin{document}

\title{A plug-and-play solution for characterizing two-way optical frequency transfer over free-space}

\author{Jingxian Ji,\authormark{1,2,*} 
Shambo Mukherjee,\authormark{1} 
Alexander Kuhl,\authormark{1} 
Sebastian Koke,\authormark{1} 
Markus Leipe,\authormark{3,4} 
Markus Rothe,\authormark{3} 
Fabian Steinlechner,\authormark{3,4} 
and Jochen Kronjäger\authormark{1}}

\address{\authormark{1}Physikalisch-Technische Bundesanstalt (PTB), Braunschweig 38116, Germany\\
\authormark{2}DLR-Institute for Satellite Geodesy and Inertial Sensing, Hannover 30167, Germany\\
\authormark{3}Fraunhofer Institute for Applied Optics and Precision Engineering (IOF), Jena 07745, Germany\\
\authormark{4}Friedrich Schiller University Jena, Institute of Applied Physics, Abbe Center of Photonics, Jena 07745, Germany}

\email{\authormark{*}jingxian.ji@ptb.de} 


\begin{abstract*} 
Optical clock networks connected by phase-coherent links offer significant potential for advancing fundamental research and diverse scientific applications. Free-space optical frequency transfer extends fiber-based connectivity to remote areas and holds the potential for global coverage via satellite links. Here we present a compact and robust portable, rack-integrated two-way free-space link characterization system. Equipped with plug-and-play capabilities, the system enables straightforward interfacing with various optical systems and facilitates quick deployment for field experiments. In this work, we achieve a fractional frequency instability of \(2.0 \times 10^{-19}\) for an averaging time of 10\,s over a 3.4\,km horizontal fully folded intra-city free-space link. Moreover, the system maintains an uptime of \SI{94}{\percent} over 15 hours, illustrating its reliability and effectiveness for high-precision optical frequency comparisons over free-space. 

\end{abstract*}

\section{Introduction}

The development of optical atomic clocks has led to unparalleled levels of accuracy, achieving fractional systematic uncertainties at the \(10^{-18}\) level and even lower~\cite{Ush2015, Hun2016, Bre2019, Zhi2023, Aep2024}. This remarkable accuracy opens up possibilities for testing fundamental theories such as general relativity~\cite{Dit2009, Alt2015} or variations of fundamental constants~\cite{Uza2011}, and a wide range of high-impact applications in geodesy~\cite{Meh2018, McG2018} and space~\cite{Cac2009}. 

These applications demand an accurate comparison of frequencies between distant optical clocks, implying the need for reliably linked optical atomic clocks. Consequently, ultra-stable phase-coherent time and frequency transfer is essential for developing global optical clock networks, requiring the capability to maintain uncertainties as low as \(10^{-18}\) over extensive distances. In the last two decades, interferometrically stabilized fiber links~\cite{ma1994, For2007} have proven their ability to transfer frequencies with uncertainties below \(10^{-19}\) over continental distances~\cite{Rau2015, Chi2015}, providing a backbone for terrestrial geodetic networks. 

However, the limited availability of fiber links necessitates alternative strategies to extend the reach and capabilities of these networks, especially in cases where fibers cannot be easily or economically deployed. This becomes particularly crucial in chronometric geodesy, where optical free-space links provide a vital alternative, enabling precise and extensive geodetic measurements in locations where traditional fiber links are impractical. Furthermore, in combination with recent developments in optical satellite communications, free-space optical links could significantly enhance global time and frequency transfer capabilities.

The most significant challenge for free-space optical links is atmospheric turbulence, as it can cause deflection and disruption of the beam path~\cite{Rob2016, And2005}. This turbulence induces variations in the air density along the transmission path, leading to refractive index changes that deform the wavefront of the optical beam, severely degrading the accuracy and reliability of frequency transfer over long distances. As such, addressing these disturbances is crucial for improving the viability of free-space optical links for high-accuracy applications such as satellite communications and global quantum key distribution (QKD). Recent advancements in the field demonstrate considerable progress in improving the uncertainty and operational range of free-space optical links for optical clock comparisons. There are two primary approaches to compare optical clocks over optical free-space links. One is optical two-way time and frequency transfer (OTWTFT) to transmit optical pulses using frequency combs~\cite{Gio2013, She2022, Cal2023}. Another approach is to use continuous-wave (CW) lasers as signal sources together with beam stabilization systems, typically employing the phase-locked loop (PLL) method for active noise cancellation (ANC)~\cite{Kan2019, Dix2021, Goz2022, mar24}. 

In this study, we analyze a CW laser-based two-way (TW) frequency transfer technique, which offers improved noise rejection and potentially reduced sensitivity to turbulence compared to the ANC schemes~\cite{Cal2014}, as both beams traverse the free-space path only once. While the two-way technique, based on post-processing of recorded data, does not support real-time delivery of ultrastable light from one location to another, it is well suited for optical clock comparisons for example in the context of chronometric geodesy. Additionally, it avoids the technical complexities associated with the ANC method. Our two-way setup, devoid of interferometric noise contributions and laser phase noises, enables the characterization of the non-reciprocity of free-space connections relevant to frequency transfer performance, which affects both techniques.

The entire setup is housed within a transportable rack optimized for off-site deployment. Its plug-and-play design, which includes a connection through a single-mode (SM) fiber patch cable, allows for straightforward interfacing with various optical systems. This versatility enables the setup to be used in-house or integrated and tested with other free-space links that employ different beam pointing stabilization strategies. Such functionality improves the adaptability of the system and simplifies its installation in diverse experimental and operational settings. Moreover, our setup enables the examination of crucial aspects and technological performance for optical clock comparisons, including the link stability and the system's portability. 

This article is organized as follows: Initially, we outline the two-way frequency transfer scheme implemented in our study in section~\ref{sec:methods}. Next, in section~\ref{sec:results}, we present the experimental results from a measurement campaign involving a 3.4\,km free-space link and discuss the performance of the setup and its primary limiting factors. Finally, we provide our conclusions in section~\ref{sec:conclusion}.

\section{Methods}
\label{sec:methods}

\begin{figure}[htbp]
\centering\includegraphics[width=\textwidth]{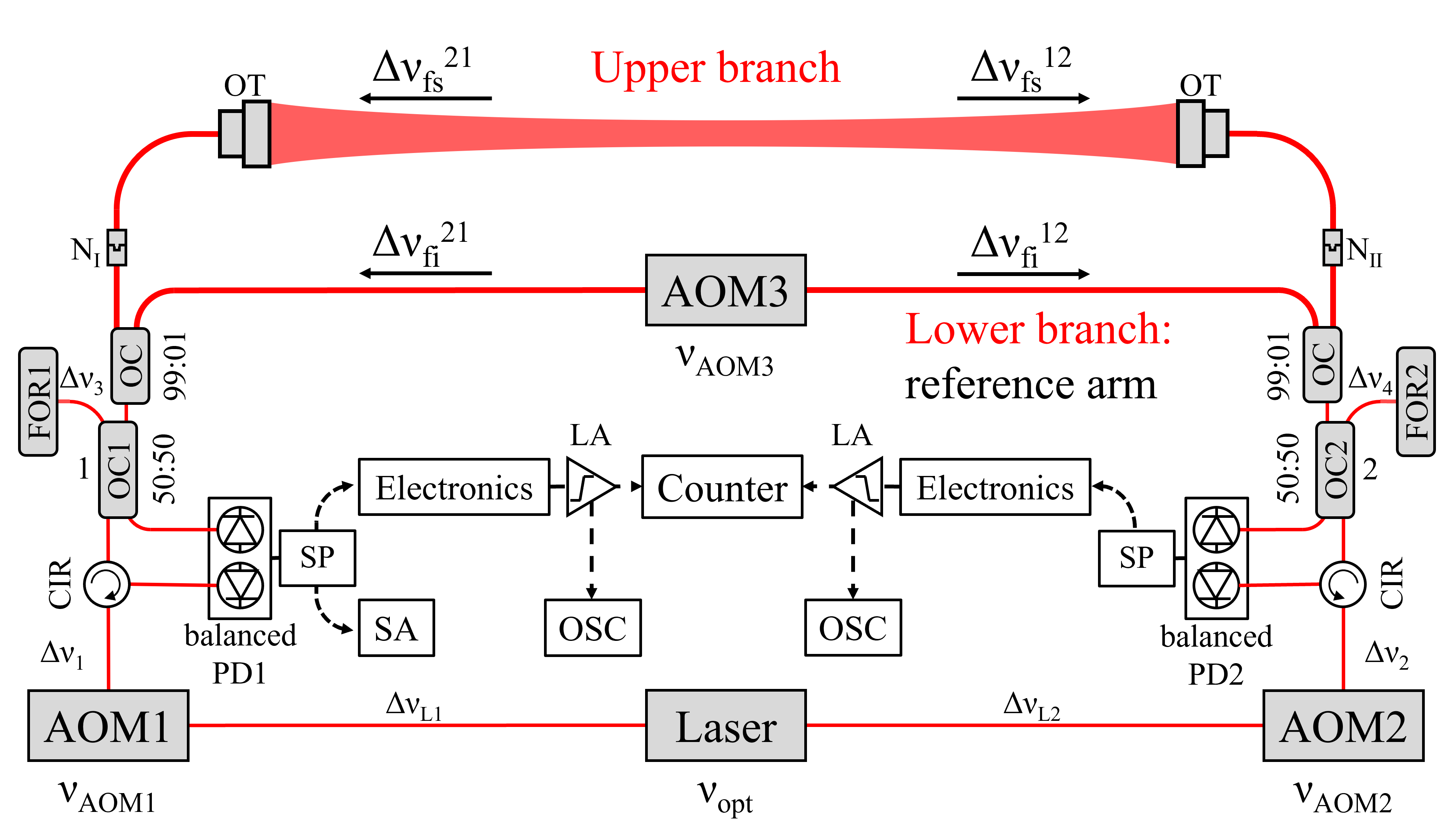}
\captionsetup{width=\textwidth}
\caption{Experimental two-way bidirectional layout for free-space link characterization. AOM, acousto-optic modulator; CIR, circulator; FOR, fiber optic retroreflector; LA, limiting amplifier; N, connection node; OC, optical coupler; OSC, oscilloscope; OT, optical terminal; PD, photodetector; SA, spectrum analyzer; SP, 1-by-4 splitter; solid red line, PM fiber; dashed black line, coaxial cable.}
\label{fig: setup_tw}
\end{figure}

The configuration of the two-way scheme for free-space link characterization is shown in Fig.~\ref{fig: setup_tw}. This configuration comprises two branches: the upper branch incorporates the free-space transmission path under test, and the lower branch serves as a reference arm, which is kept short. The nodes labeled $\mathrm{N}_{\text{I}}$ and $\mathrm{N}_{\text{II}}$ indicate the connection points to the free-space transmission path, allowing flexible integration of different link configurations including fully folded links for testing as described below. Additionally, this setup can accommodate a partially folded link where the transmitter and receiver are closely co-located, as detailed in~\cite{Ji2023}. 

A 20\,dBm optical signal generated by a 1542\,nm fiber laser (Koheras ADJUSTIK E15) with low phase noise is split, and on each side, its frequency is shifted with acousto-optic modulators (AOMs), with frequency shifts of $\nu_{\text{AOM1}}$ and $\nu_{\text{AOM2}}$. These two AOMs practically simulate the scenario of separate lasers, as would be employed in real-world point-to-point optical frequency comparisons. The signal $\nu_{1}=\nu_\text{{opt}}+\nu_{\text{AOM1}}$ is directed into two branches from left to right, while the signal $\nu_{2}=\nu_\text{{opt}}+\nu_{\text{AOM2}}$ from right to left. On each side of the setup, a Michelson interferometer configuration is implemented, employing fiber optic retroreflectors (FOR) to reflect portions of $\nu_{1}$ and $\nu_{2}$ light. These reflected signals act as local oscillators or interferometric beat partners, effectively creating a double two-way link with the same phase reference planes. The entire setup is interconnected using polarization-maintaining (PM) fiber to ensure that the polarization state of the signals remains constant despite environmental variations. Two circulators (CIR), positioned after AOM1 and AOM2, are used for balanced photodetection to enhance the signal-to-noise ratio (SNR) and efficiently suppress some unwanted beat notes. The insertion loss introduced by each circulator is compensated by adding an equivalent attenuation to the other arm of the balanced photodetector. The entire system, encompassing the two-way optical setup and the electronics required for signal processing, is integrated into a transportable rack. 

This configuration provides three beat notes on each of the two balanced photodetectors PD1 and PD2 at each end: A, the beat note between the local oscillator and remote laser through the lower branch, which is the reference arm with an extra frequency shift of $\nu_{\text{AOM3}}$; B, the beat note between the local oscillator and remote laser through the upper branch incl. the free-space link and any additional frequency shifts; and C, the beat note between the remote laser respectively through the upper and lower branches. The extra frequency shifts are essential for differentiating the useful beat notes from parasitic back reflections. The frequencies of the beat notes detected by the photodetectors PD1 and PD2 can be written as follows:

\begin{equation}
    \begin{aligned}
        \text{PD1}_{A} &= \nu_{2} + \Delta\nu_{fi}^{21} - \nu_{1} + \text{n}_1,\label{eq:beat notes}\\
        \text{PD1}_{B} &= \nu_{2} + \Delta\nu_{fs}^{21} - \nu_{1}+ \text{n}_1,\\
        \text{PD1}_{C} &= \nu_{2} + \Delta\nu_{fs}^{21} - (\nu_{2} + \Delta\nu_{fi}^{21}),\\
        \text{PD2}_{A} &= \nu_{1} + \Delta\nu_{fi}^{12} - \nu_{2}+ \text{n}_2,\\
        \text{PD2}_{B} &= \nu_{1} + \Delta\nu_{fs}^{12} - \nu_{2}+ \text{n}_2,\\
        \text{PD2}_{C} &= \nu_{1} + \Delta\nu_{fs}^{12} - (\nu_{1} + \Delta\nu_{fi}^{12}).
    \end{aligned}
\end{equation}

where $\Delta\nu_{fi}$ and $\Delta\nu_{fs}$ are the frequency noise contributions (including shifts) introduced by the short reference arm and free-space link, respectively, as the light propagates from left to right ($\Delta\nu_{fi}^{12}$, $\Delta\nu_{fs}^{12}$) and from right to left ($\Delta\nu_{fi}^{21}$, $\Delta\nu_{fs}^{21}$). The terms $\text{n}_1$ and $\text{n}_2$ denote the interferometric noise contributions, which are described in detail in Appendix~\ref{sec:IF noise}.

After photodetection, the A and B beat notes are frequency converted, band-pass filtered and amplified in a three-stage heterodyne receiver. A limiting amplifier, based on the integrated circuit AD8306 from Analog Devices, is used to produce a stable output. The signal processing chain is designed for a high dynamic range, with strong suppression of out-of-band signals and negligible added noise. The full signal bandwidth of the electronics is approximately 3.5\,MHz with a dynamic range of approximately 50\,dB. A block diagram of the three-stage heterodyne receiver is shown in Appendix~\ref{sec:electronics}. All beat signals are then simultaneously counted using deadtime-free frequency counters in $\Lambda$-mode~\cite{ben2015}.

The radio frequency (RF) power levels of beat notes $\text{PD1}_{A}$ and $\text{PD1}_{B}$ are monitored continuously on a spectrum analyzer with a 1\,s resolution. Concurrently, the limiting amplifier, featured with a logarithmic voltage readout, efficiently demodulates the RF power. This voltage is then recorded using an oscilloscope for high-resolution measurements.

Combining the signals labeled as A and B from two photodetectors at each end in Equation~\eqref{eq:beat notes} in post-processing, one can obtain
\begin{equation}
    (\text{PD1}_{A} - \text{PD1}_{B}) -  (\text{PD2}_{A} - \text{PD2}_{B}) =  (\Delta\nu_{fi}^{21} - \Delta\nu_{fi}^{12}) - (\Delta\nu_{fs}^{21}-\Delta\nu_{fs}^{12}).\label{eq:a+b}
\end{equation}

In addition, an equivalent result can be achieved by combining the C beat notes detected on photodetectors PD1 and PD2
\begin{equation}
    \text{PD2}_{C} - \text{PD1}_{C} = (\Delta\nu_{fi}^{21} - \Delta\nu_{fi}^{12}) - (\Delta\nu_{fs}^{21}-\Delta\nu_{fs}^{12}). \label{eq:c}\\
\end{equation}

In Equations~\eqref{eq:a+b} and~\eqref{eq:c}, all interferometric noise terms, including those from local fiber interconnections and laser phase noise are common mode, thus eliminated through data post-processing. The term $\Delta\nu_{fi}^{21}-\Delta\nu_{fi}^{12}$ is included in the noise floor measurement by short-cutting the free-space connection. It has been thoroughly characterized and found to be negligible. Consequently, the expressions on the left sides of Equations~\eqref{eq:a+b} and~\eqref{eq:c} indicate the (apparent) residual non-reciprocal noise arising from the free-space link, which serves as the key variable for frequency comparisons. Conversely, in an actual clock comparison the frequency difference $\nu_1 -\nu_2$ can be obtained from the combination

\begin{equation}
    \text{PD1}_{B} - \text{PD2}_{B} = 2\times(\nu_2 - \nu_1) + (\Delta\nu_{fs}^{21}-\Delta\nu_{fs}^{12}) + (\text{n}_1-\text{n}_2). \label{eq:b}\\
\end{equation}

Note that Equation~\eqref{eq:b}, in contrast to Equations~\eqref{eq:a+b} and~\eqref{eq:c}, contains interferometric noise terms. 

To assess the performance of the TW free-space link characterization setup, we conducted several field tests. After the first test~\cite{Ji2023} on the PTB campus, where the maximum distance was constrained to 600\,m, we proceeded with a free-space testbed~\cite{Goy2021} in Jena, Germany. 

This 1.7\,km intra-city link, initially designed as QKD infrastructure, contains two mobile optical terminals situated at the Fraunhofer Institute for Applied Optics and Precision Engineering (IOF) building and Stadtwerke (STW) Jena. Each optical terminal is equipped with a beacon laser operating at wavelengths of 1064\,nm and 980\,nm respectively, alongside an active 4D beam stabilization system. The system uses two closed-loop controlled tip-tilt mirrors on each side to achieve single-mode fiber coupling. A pair of position-sensitive devices measures the position and angle of the beacon laser beam, effectively stabilizing the payload beam at 1550\,nm. The control loop allows for quick adjustments to compensate for atmospheric turbulence impacts such as tip/tilt wavefront aberrations and optical beam wander. Atmospheric turbulence is monitored using an independent scintillometer (Scintec BLS900), whose transmitter and receiver are also located at IOF and STW buildings, respectively. This instrument provides measurements of the refractive index structure parameter $C_n^2$, which quantifies the strength of turbulence-induced refractive index fluctuations in the atmosphere~\cite{And2005}.

\begin{figure}[htbp]
\centering\includegraphics[width=\textwidth]{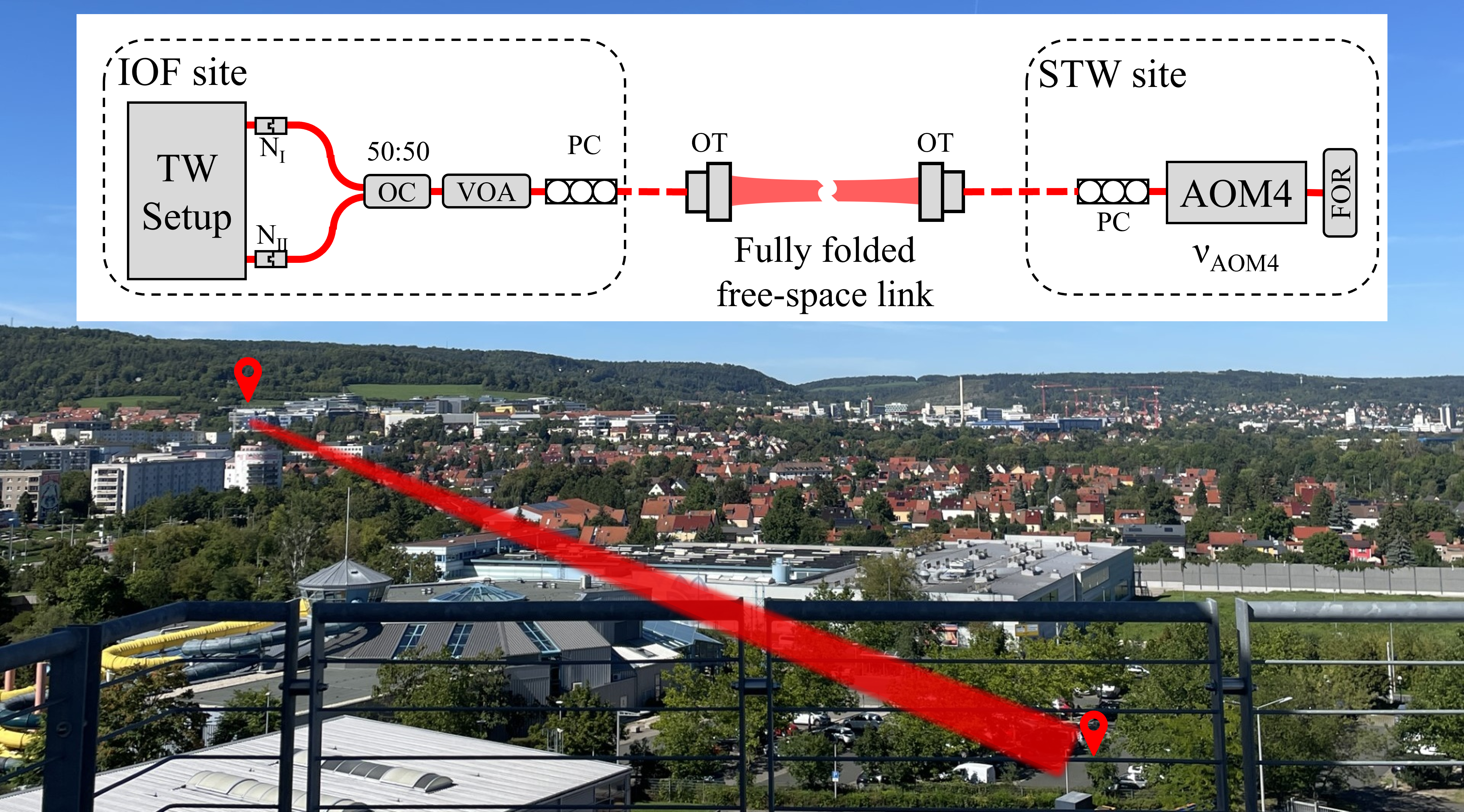}
\captionsetup{width=\textwidth}
\caption{Overview of the experimental setup with the 1.7\,km free-space link in Jena, Germany. FOR, fiber optic retroreflector; PC, polarization controller; VOA, variable optical attenuator; dashed red line, SM fiber.}
\label{fig: setup_jena}
\end{figure}

The optical terminals utilize an unobscured, afocal four-mirror metal telescope with a 200\,mm aperture and a magnification of $\times20$. The nominal $1/e^{2}$ beam diameter is 64\,mm. The terminals provide full plug-and-play capabilities for the payload signal at 1550\,nm, ensuring straightforward integration with our setup. As shown in Fig.~\ref{fig: setup_jena}, our TW setup was then connected to the local optical terminal located in the laboratory of the IOF building via an optical coupler, with one side linking nodes $\mathrm{N}_{\text{I}}$ and $\mathrm{N}_{\text{II}}$ and the other side connected to the terminal through a 5\,m SM fiber. The AOMs used in this campaign operated at frequency shifts of $\nu_{\text{AOM1}} = +37\,\mathrm{MHz}$, $\nu_{\text{AOM2}} = +47\,\mathrm{MHz}$, and $\nu_{\text{AOM3}} = -71\,\mathrm{MHz}$. The total loss of the setup from the laser to the input of the optical terminal was 18.5\,dB, resulting in a total optical power of 1.5\,dBm entering the optical terminal.

An FOR with an AOM operating at a frequency shift of $-38.5\,\mathrm{MHz}$ was located at the STW site, creating the fully folded free-space link of 3.4\,km, i.e., a link where forward and backward beam path exactly match. The light path involves the light traveling both ways in the upper branch to pass simultaneously through a given element of the free-space path twice, ensuring that the phase delay experienced is identical in both directions. Hence, the delay-induced non-reciprocity contributions to $\Delta\nu_{fs}^{12}$ and $\Delta\nu_{fs}^{21}$ exactly cancel out. Thus, under ideal experimental conditions, delay noise~\cite{New2007} should not be present, allowing the measurements to reveal the technical limitations of the setup. Frequency-dependent phase shifts may still lead to a detectable non-reciprocity, as the light beams entering through $\mathrm{N}_{\text{I}}$ and $\mathrm{N}_{\text{II}}$ have different frequencies due to the AOM shifts. However, no evidence of such effects was observed in the measurements. Disregarding dispersion~\cite{Lop2010}, we estimate a residual non-reciprocity of approximately $5\times10^{-8}$ of the free-running Doppler shift as a result of AOM shifts (see Appendix~\ref{sec:frequency} for details), which is well below our observations.
 At each side, a polarization controller is placed at the transition from PM to SM fiber to ensure proper alignment of the polarization. Although the link is fully folded for frequency transfer, the system effectively replicates the beam stabilization of a true point-to-point free-space link.

\section{Results}
\label{sec:results}

In this section, we present the results from the measurement campaign conducted in October 2023 over the course of 3 weeks in Jena, Germany. During the measurement campaign, the powers of the C beat notes were too low for effective detection. Therefore, Equation~\eqref{eq:a+b} was used to evaluate our data.

\begin{figure}[htbp]
\centering\includegraphics[width=\textwidth]{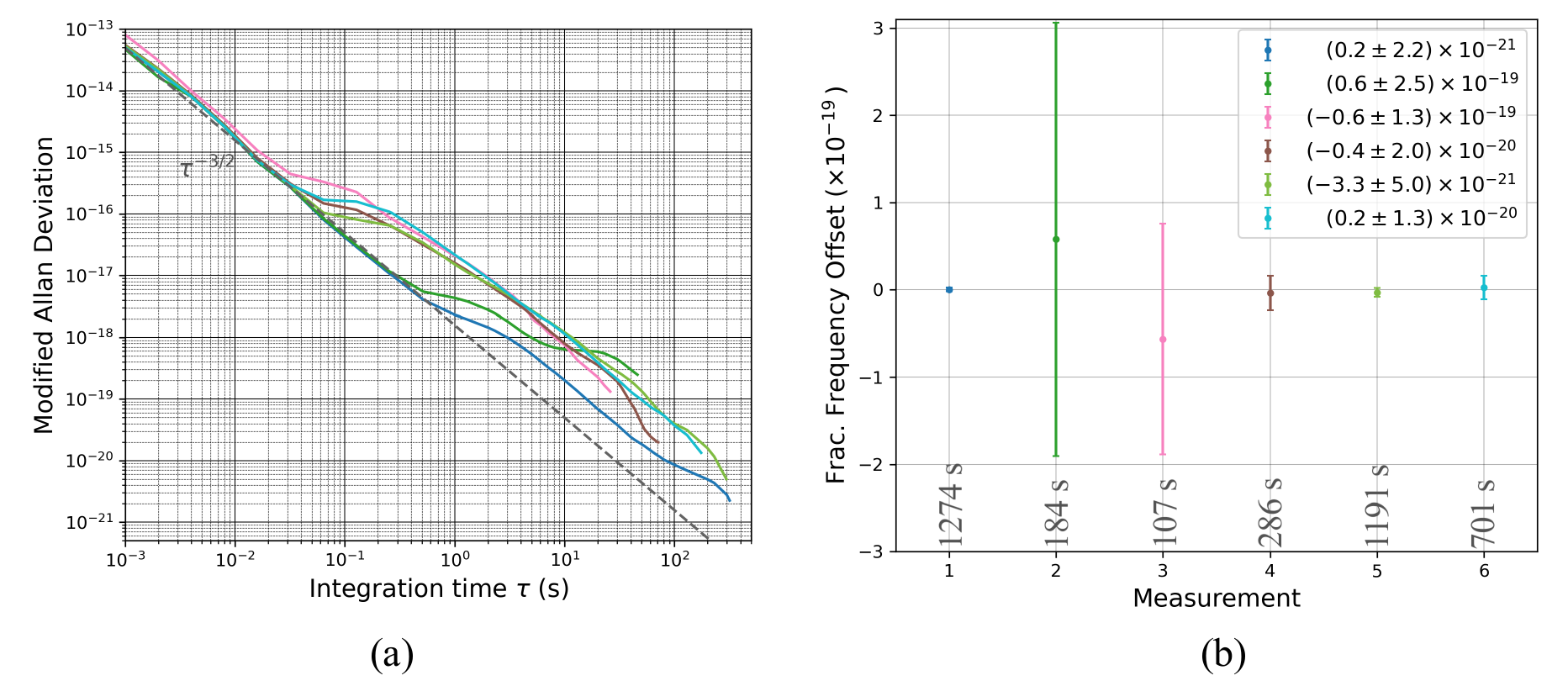}
\captionsetup{width=\textwidth}
\caption{(a) Scatter of the modified Allan deviation of longest continuous measurement runs on different days. Different runs are coded by distinct colors. (b) The corresponding $\Lambda$-averaged fractional frequency offsets over all measurement runs, with error bars determined by the last value of the corresponding modADEV at 1/4 measurement time. The labels indicate the duration of each measurement run.}
\label{fig: 1ms_all}
\end{figure}

In Fig.~\ref{fig: 1ms_all}(a), we investigate how the observed instability varies from measurement to measurement and show the variations in terms of modified Allan deviation (MDEV) of the free-space link across the longest measurement runs with 1\,ms resolution time. All data presented here represent continuous periods free of cycle slips. The frequency transfer accuracy of the corresponding measurement runs is summarized in Fig.~\ref{fig: 1ms_all}(b). 

The short-term MDEV, which averages down proportionally to $\tau^{-3/2}$, indicates that white phase noise, primarily due to shot noise, is the dominant noise at these timescales. Notably, the short-term instability remains consistent across all measurement runs. This aspect will be discussed in further detail later. However, the long-term instability, occurring at integration times of 0.2 seconds and above, demonstrates variability across different measurements. 

Among these, measurement 1 exhibits the lowest instability. For this specific measurement, the modified Allan deviation is initially $2.3\times10^{-18}$ at an integration time $\tau$~=~1\,s. As the integration time increases, the MDEV reaches an instability of $2.0\times10^{-19}$ at 10\,s, further reducing to $8.5\times10^{-21}$ for an averaging time of $\tau$~=~100\,s. 

For measurements 3 to 6, the MDEV plots exhibit a notable bump around 0.1\,seconds, with the magnitude approximately ten times higher. These bumps may be related to modifications in the telescope alignment, particularly after the optical terminal required realignment due to a malfunction in its control system between measurements 2 and 3. This adjustment potentially altered the beam pointing, impacting fiber coupling efficiency and thereby the amplitude characteristics of the received signals. The exact mechanisms behind these anomalies are not yet clear and require further investigation.

Overall, at a 10\,s integration time, the MDEV varies between $2.0\times10^{-19}$ and $1.1\times10^{-18}$. This performance surpasses that of state-of-the-art clock accuracy, indicating that even a sufficient number of very short ($\gtrsim 10\,\mathrm{s}$) cycle-slip free intervals are sufficient for accurate clock comparisons.

Figure~\ref{fig: 1ms_all}(b) demonstrates that all fractional frequency offsets are scattered within the low \(10^{-19}\) range. The observed offsets are compatible with zero within $1\sigma$, where $\sigma$ represents the statistical uncertainty of the average~\cite{ben2015}. The apparently large errors observed in measurements 2 and 3 are primarily due to their shorter duration compared to other measurements, which leads to greater deviations and instability.

\begin{figure}[htbp]
\centering\includegraphics[width=\textwidth]{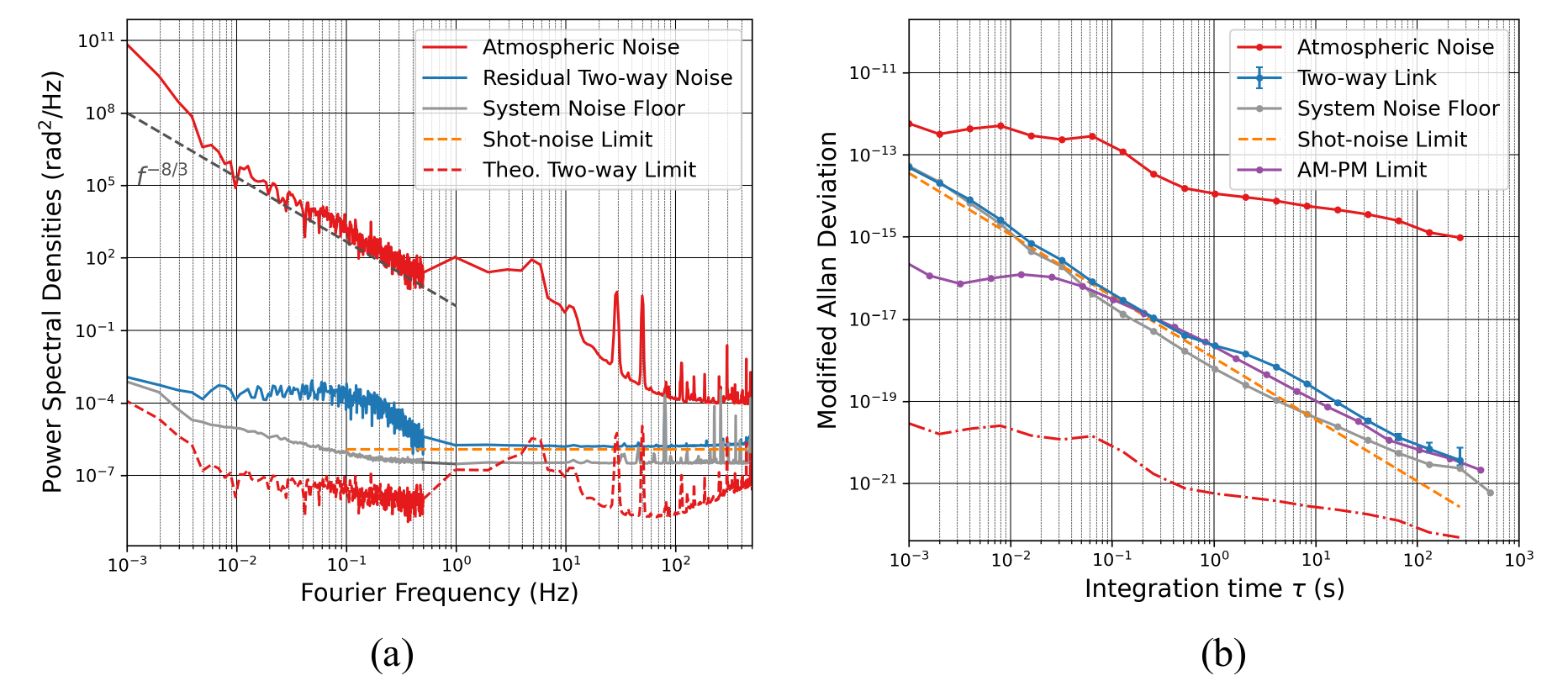}
\captionsetup{width=\textwidth}
\caption{Phase and frequency stability of the 3.4\,km folded free-space link. (a) Power spectral density (PSD) of the phase noise. Blue curve, the residual two-way free-space link noise; grey curve, the noise floor of the setup with short fiber connection; red curve, free-running phase noise; dashed red curve, calculated link delay noise, evaluated from solid red curve; dashed orange curve, calculated shot-noise evaluation of the setup with the beat amplitude of \( -32\,\text{dBm} \). (b) Fractional frequency stability as modified Allan deviation. Purple curve, AM-PM evaluation from beat amplitude; red dot-dash curve: estimated frequency-dependent residual non-reciprocity, calculated from the red curve with a factor of $5\times10^{-8}$.}
    \label{fig: noise limit}
\end{figure}

For a more detailed investigation into the underlying noise sources, measurement 1 has been selected as the primary focus due to its minimal instability. This measurement, lasting 1274\,s or 21\,minutes, represents the longest cycle-slip free interval observed across all 6 measurements presented here.

Figure~\ref{fig: noise limit}(a) shows the results of the phase noise power spectral density (PSD) over the 3.4\,km free-space link. The blue curve represents the phase noise of the two-way compensated free-space link, while the gray curve indicates the system noise floor, which is determined using a short fiber in place of the free-space link. The red curve plots the free-running link noise, capturing the atmospheric noise throughout the entire path. When comparing the red and blue curves at 1\,Hz, a reduction of eight orders of magnitude in phase noise PSD is observed for two-way compensated signals compared to the turbulence induced phase noise, reducing it to $3.4 \times 10^{-7}\,\text{rad}^{2}/\text{Hz}$.

The theoretical residual phase noise, as defined by the delay limit for a hypothetical straight, unfolded point-to-point link assuming spatially uncorrelated noise, which represents a realistic scenario for optical clock comparisons, can be estimated~\cite{Ste15} using data from the simultaneous measurement of the free-running link noise. For a point-to-point link with the total length of 3.4\,km and noise conditions as observed on our fully folded link, the calculated noise limit is depicted as a red dashed curve in Fig.~\ref{fig: noise limit}(a). A comparison between the experimentally observed residual two-way noise and this theoretical residual noise reveals that the non-reciprocity observed in the free-space link exceeds the theoretical predictions. Even in the point-to-point setup, the delay limit would not be observable, indicating that other factors must be affecting the observed non-reciprocity.

Figure~\ref{fig: noise limit}(b) presents the fractional frequency instability, illustrating both the free-space link noise and the system noise floor. The system noise floor measurement indicates a $\Lambda$-averaged offset of $1.25\times10^{-21}$, signifying zero compatible operation during this measurement. Initially, the MDEV for both the free-space link and system noise floor show consistent short-term instability. However, at longer integration time, the two-way link non-reciprocity exhibits increased instability. To identify the specific contributing factors, we have included the analysis of the expected amplitude-to-phase noise conversion (AM-PM) and shot noise from the photodetector. 

In our analysis of potential noise sources within the setup, the AM-PM effect is identified as a significant factor. This effect primarily originates from the electronic components within the setup. Amplitude modulation in the free-space link, i.e. fluctuations of the received optical power, is caused by atmospheric turbulence that the beam stabilization system cannot fully mitigate.  Such modulations, in turn, impact the amplitude stability of the received RF signal, subsequently leading to phase modulation due to the nonlinear response of the electronic components involved in signal processing. 

\begin{figure}[htbp]
\centering\includegraphics[width=\textwidth]{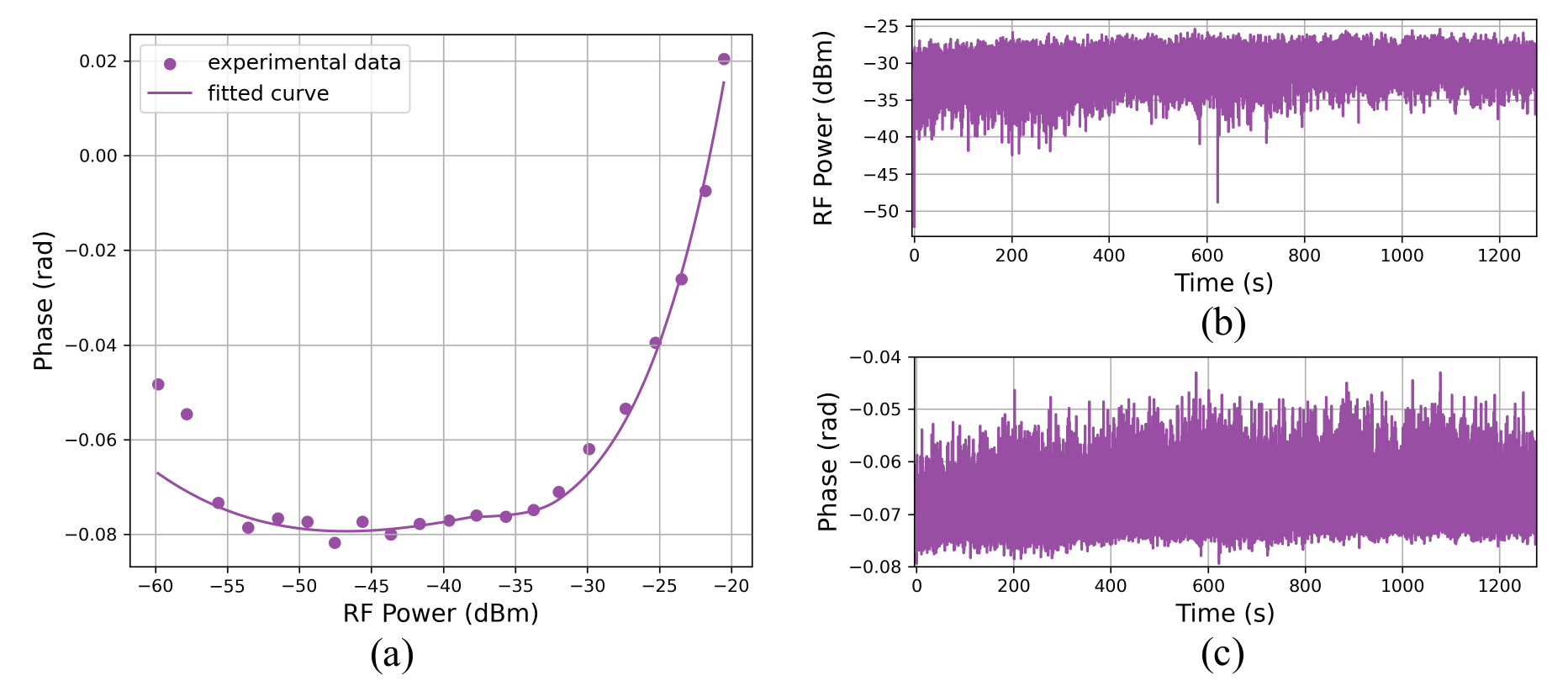}
\captionsetup{width=\textwidth}
\caption{(a) Phase response to RF power, (b) received RF signal power sampled at 5\,kHz, (c) calculated phase response.}
\label{fig: am_pm}
\end{figure}

To investigate the relationship between amplitude modulation and phase modulation, an additional experiment was conducted in the laboratory, in which the free-space link in the setup shown in Fig.~\ref{fig: setup_jena} was replaced with a short fiber. In this experiment, we adjusted the power of the detected RF signal from \(-60\) to \( -20\,\text{dBm} \) in 2\,dB increments by adjusting the variable optical attenuator (VOA) in the setup, while the data was recorded continuously to preserve phase-shift information. The experimental data points are shown as purple dots in Fig.~\ref{fig: am_pm}(a), demonstrating the phase shift at various RF power levels. The curve fitting the data is divided into two segments: for RF power levels lower than \( -38\,\text{dBm} \), the relationship is modeled with a third-degree polynomial fit. For RF power levels above \( -38\,\text{dBm} \), the relationship is modeled by an exponential fit, capturing the sharp increase in phase shift as power increases. A total phase shift of 0.1\,radians is observed over a 40\,dB dynamic range. The detailed functional forms are given in Appendix~\ref{sec:AM-PM}.

Using this relationship, we transformed the measured amplitude fluctuations within the same interval as measurement 1, sampled at 5 kHz using an oscilloscope, into phase data. These transformations are presented in Figs.~\ref{fig: am_pm}(b) and~\ref{fig: am_pm}(c). The modified Allan deviation calculated from this phase data is depicted in purple in Figure~\ref{fig: noise limit}(b). This plot reveals that for integration times above 0.2 seconds (below 2.5\,Hz in PSD), the performance is limited by AM-PM conversion.

This observation explains the variations in long-term instability observed across different measurements, as illustrated in Figure~\ref{fig: 1ms_all}(a), which are directly influenced by daily fluctuations in atmospheric turbulence. 

At shorter time scales, the instability is determined by white phase noise, which can be quantified by the inverse of the signal-to-noise ratio. In our experiment, the photodetector is operated in the shot-noise limited regime due to the use of heterodyne detection, ensuring that local oscillator shot noise dominates over all other noise sources. The achievable SNR~\cite{kin1978} can be determined by the received optical power
\begin{equation}
    \mathrm{SNR}_{\text{sn}} = 10 \log_{10} \left( \frac{\mathcal{R} P_S}{q} \right) \label{eq:SNR}\\
\end{equation}
where $q$ is the elementary charge ($1.602\times 10^{-19}$ C), $\mathcal{R}$ is the responsivity of the photodetector (0.95 A/W),  $P_S$ is the optical power of the signal. In $\mathrm{dBm}$, the relationship is given by
\begin{equation}
\mathrm{SNR}_{\mathrm{sn}} = P_{S} + 157.7.
\label{eq:SNR_dBm}\\
\end{equation}

Although $P_S$ is not directly measured in our setup, it can be estimated from the measured RF beat power $P_{\mathrm{RF}}$ by using the effective transimpedance gain ($1.25\,\text{k}\Omega$ in our case, accounting for the $50\,\Omega$ output impedance and a subsequent 1-by-4 splitter), and the measured optical power of the local oscillator beam $P_{\mathrm{LO}}$. 

Based on standard heterodyne detection principles, the relationship in dBm is:

\begin{equation}
    P_S = P_{\mathrm{RF}} - P_{\mathrm{LO}} - 17.5. \label{eq:P_RF}
\end{equation}

In our setup, the measured local oscillator power was $1.1\,\mathrm{dBm}$ for each photodetector, corresponding to a total of $P_{\mathrm{LO}}=4.1\,\mathrm{dBm}$. Thus, the achievable $\mathrm{SNR}_{\text{sn}}$ can be determined from the measured beat power $P_{\mathrm{RF}}$ using~\eqref{eq:SNR_dBm} and~\eqref{eq:P_RF}.

During the measurement 1, the RF power of $\text{PD1}_{A}$ remained consistently stable with an average of \( -32.3\,\text{dBm} \), corresponding to a shot-noise limited SNR of \( 103.8\,\text{dB} \). In contrast, the RF power of $\text{PD1}_{B}$ fluctuated, with an average of \( -31.2\,\text{dBm} \), as illustrated in Fig.~\ref{fig: am_pm}(b). Assuming that the RF power levels on both photodetectors PD1 and PD2 are the same, the phase noise for the combined A and B signals after post-processing is estimated to increase by +6.0\,dB relative to that of $\text{PD1}_{A}$. The detection electronics have an effective filter bandwidth of 3.5\,MHz, which results in a \(+35.4\,\text{dB}\) increase in phase noise due to aliasing by the frequency counter sampling at 1\,kHz. Consequently, the white phase noise $S_{\varphi}(f)$ is then expected at a level of $1.2 \times 10^{-6} \text{ rad}^2/\text{Hz}$ as illustrated in Fig.~\ref{fig: noise limit}(a). Following the equation in ~\cite{rub2023}, the modified Allan deviation is computed and shown in orange in Fig.~\ref{fig: noise limit}(b). It matches the observed frequency transfer instability well, indicating that the short-term instability of the non-reciprocity is essentially limited by the shot noise.

\begin{figure}[htbp]
\centering\includegraphics[width=\textwidth]{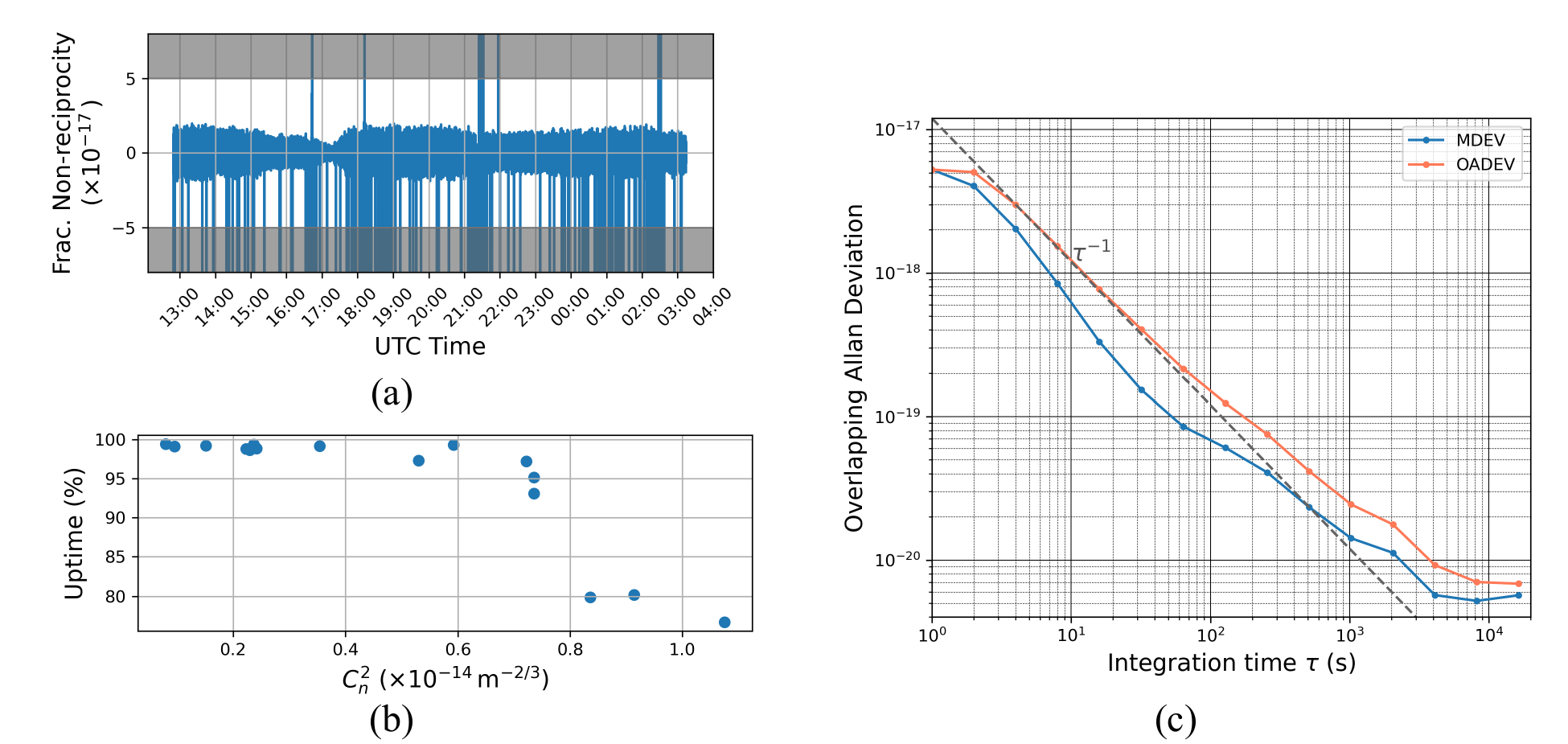}
\captionsetup{width=\textwidth}
\caption{Results of a measurement run, spanning from 2023-10-12 at 12:49 to 2023-10-13 03:15 with 1\,s resolution in time. (a) Measured fractional non-reciprocity. Data points exceeding the thresholds of $\pm 5\times 10^{-17}$ are considered invalid. (b) Hourly variation of the system uptime with atmospheric turbulence strength. (c) OADEV and MDEV for the valid fractional non-reciprocity, calculated from concatenated cycle-slip free intervals.}
\label{fig: uptime}
\end{figure}

To evaluate the feasibility of this free-space link for optical clock comparisons, we present time series data from a measurement that commenced on the same day as measurement 1 with 1\,s resolution in time. This time span allows for the observation of variations during both daylight and nighttime hours and at different levels of turbulence. 

The system uptime is calculated for each full hour using the ratio $N_{\text{valid}}/N_{\text{total}}$, where $N_{\text{valid}}$ is the number of valid data points and $N_{\text{total}}=3600$ is the total duration of the measurement in seconds. Invalid data points are defined as those where the measured fractional non-reciprocity (1\,s $\Lambda$ average) exceeds  $\pm 5\times 10^{-17}$, as illustrated in Fig.~\ref{fig: uptime}(a). This threshold was chosen to be significantly larger than the typical peak-to-peak non-reciprocity, yet smaller than one cycle slip. It is predominantly set to identify and exclude anomalies caused by deep fades, which occur when the received power drops below the lower end of the dynamic range of our detection electronics (roughly \( -60\,\text{dBm} \)). Considering that these deep fades typically last for only a few milliseconds, they are too transient to be detected by 1\,Hz sampling.

Figure~\ref{fig: uptime}(b) illustrates the relationship between the turbulence strength, represented by $C_n^2$, and the system uptime percentage. Over a measurement period exceeding 15 hours, 12 hours exhibit uptime $\geq$ \SI{90}{\percent}. As $C_n^2$ increases, indicating stronger atmospheric turbulence, the system uptime correspondingly decreases. Initially, for lower $C_n^2$ values, the uptime remains nearly at \SI{100}{\percent}. However, as $C_n^2$ increases towards $0.8\times10^{-14}\,\text{m}^{-2/3}$, the uptime declines markedly, dropping below \SI{80}{\percent}. This trend shows the impact of turbulence on system performance, where higher turbulence levels correlate with reduced uptime. 

Despite this, the overall uptime for the link during the entire measurement period was \SI{94}{\percent}. This level of operational reliability is essential for optical clock comparisons, as optical clocks require sufficient averaging time to achieve the anticipated relative frequency uncertainty. 

As shown in Fig.~\ref{fig: uptime}(c), the overlapping Allan deviation (OADEV) reached an instability of $6.5\times 10^{-21}$ for an averaging time of $\tau = 12\,\mathrm{ks}$. The $\Pi$-averaged offset was $-1.4 \times 10^{-21}$, which is compatible with zero. Remarkably, the statistical uncertainty achieved is three orders of magnitude below the typical statistical uncertainty observed in optical clock comparisons over the same period~\cite{Gro2024}. This performance demonstrates the system's robustness against potential disruptions and ensures consistent data accuracy throughout the duration of the measurement.

\section{Conclusion and Outlook}
\label{sec:conclusion}

This work has rigorously evaluated a novel two-way optical frequency transfer system tailored for evaluating free-space links for the purpose of comparing optical atomic clocks. Extensive characterization has confirmed an instability of $10^{-20}$ after 100~s of integration, demonstrating its potential for robust and precise clock comparisons. Adaptations for minimizing interferometric noise contributions in genuine point-to-point two-way free-space links for optical clock comparisons have been well-established for interferometric fiber links and are straightforward to implement~\cite{Aka2020, Kuh2024}. Furthermore, the study identifies the limiting factors in the setup. Future work will aim to identify the dominant components contributing to the AM-PM effect and to mitigate this impact. 

The setup described in this study is both compact and transportable, embodying a plug-and-play design that seamlessly integrates with existing free-space optical infrastructure. This portability and ease of deployment make it highly suitable for a variety of field experiments, accommodating variable conditions. The adaptability of the system further extends the capacity for chronometric leveling, offering the ability to resolve height differences at the centimeter level. This capability is particularly valuable in geodesy, where the flexibility of free-space connections supports application scenarios not accessible with buried fiber connections.

In the long term, our research is dedicated to enabling high-accuracy clock comparisons over distances of up to several tens of kilometers. Achieving this goal presents several technical challenges. Increased propagation distance results in greater atmospheric attenuation and beam divergence, leading to lower received optical power and reduced SNR. To detect weak incoming signals under these conditions, the signal bandwidth may need to be reduced. Moreover, higher phase noise and increased propagation delay lead to a disproportionately higher level of residual noise, exhibiting cubic scaling with the link length~\cite{Ste15},  which may complicate cycle slip detection, necessitating the development or adaptation of advanced data processing techniques, such as those used in long-haul optical fiber links. Received power fluctuations due to atmospheric turbulence will also be worse, reinforcing the need for electronics with reduced AM-PM conversion. In parallel, we aim to further enhance the accuracy and operational capabilities of our system through the implementation of high-precision pointing and tracking systems capable of compensating for turbulence-induced beam wander and mechanical instabilities, enabling more autonomous and efficient field deployments for optical frequency comparisons over extended distances.

\appendix
\makeatletter
\renewcommand{\thesection}{\Alph{section}}
\renewcommand{\@seccntformat}[1]{Appendix~\csname the#1\endcsname:\quad}
\makeatother

\section{Interferometric noise}
\label{sec:IF noise}

The additional interferometric noise terms are defined as

\begin{equation}
    \text{n}_1 =  \Delta\nu_{L2} +\Delta\nu_{2}- \Delta\nu_{L1} -\Delta\nu_{1} - 2\Delta\nu_{3},\label{eq:n1}
\end{equation}

\begin{equation}
    \text{n}_2 =  \Delta\nu_{L1} +\Delta\nu_{1}- \Delta\nu_{L2} -\Delta\nu_{2}- 2\Delta\nu_{4}.\label{eq:n2}
\end{equation}
where $\Delta\nu_{L1}$ and $\Delta\nu_{L2}$ denote the noise contributions from the laser up to AOM1 and AOM2, respectively.  
$\Delta\nu_{1}$ and $\Delta\nu_{2}$ are the noise contributions from AOM1 to OC1 and from AOM2 to OC2, respectively.  
$\Delta\nu_{3}$ and $\Delta\nu_{4}$ represent the noise contributions from OC1 to FOR1 and from OC2 to FOR2, respectively. The corresponding optical paths are indicated in Fig.~\ref{fig: setup_tw}. All other optical paths are either common-mode (i.e. common to both light waves generating a beat note), or are part of the two-way compensated path.

\section{Block diagram of the three-stage heterodyne receiver}
\label{sec:electronics}

\begin{figure}[htbp]
\centering\includegraphics[width=\textwidth]{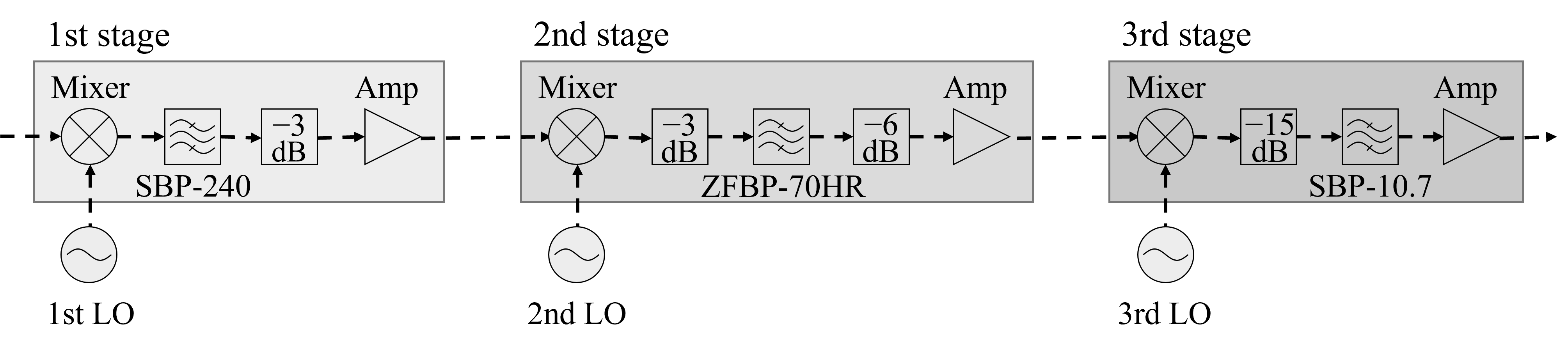}
\captionsetup{width=\textwidth}
\caption{Block diagram of the three-stage heterodyne receiver. Amp, amplifier ZFL-500; Mixer: ZFM-2; LO, local oscillator.}
\label{fig: electronics}
\end{figure}

Figure~\ref{fig: electronics} presents a schematic of the electronics block referenced in Fig.~\ref{fig: setup_tw}. It represents a self-built three-stage heterodyne receiver, which follows a conventional RF signal processing approach. All components are from Mini-Circuits, and the specific model numbers are indicated in the figure. The input signal is mixed to 240\,MHz, allowing flexible selection of the input beat note frequency. It is then down-converted to 70\,MHz and finally to 10.7\,MHz. Carefully selected band-pass filters in each stage ensure that signals spaced 5\,MHz or more are strongly suppressed, preventing unwanted signals and noise from contributing to the phase measurement.

\section{Frequency-dependent phase shifts}
\label{sec:frequency}

We consider the Doppler shift experienced by light waves travelling from $\mathrm{N}_{\text{I}}$ to $\mathrm{N}_{\text{II}}$ and vice versa via the reflector FOR at the remote site. With $\tau$ the one-way delay from $\mathrm{N}_{\text{I}}$ (or $\mathrm{N}_{\text{II}}$) to FOR, the Doppler shift from $\mathrm{N}_{\text{I}}$ to $\mathrm{N}_{\text{II}}$ is  

\begin{equation}
\Delta\nu_{\text{Doppler}}^{12} = -(\nu_{\text{opt}}+\nu_{\text{AOM1}})\frac{d\tau}{dt}-(\nu_{\text{opt}}+\nu_{\text{AOM1}}+2\nu_{\text{AOM4}})\frac{d\tau}{dt},
\end{equation}

The Doppler shift in the opposite direction $\mathrm{N}_{\text{II}}$ to $\mathrm{N}_{\text{I}}$ is  

\begin{equation}
\Delta\nu_{\text{Doppler}}^{21} = -(\nu_{\text{opt}}+\nu_{\text{AOM2}})\frac{d\tau}{dt}-(\nu_{\text{opt}}+\nu_{\text{AOM2}}+2\nu_{\text{AOM4}})\frac{d\tau}{dt},
\end{equation}

The resulting non-reciprocity as a fraction of the absolute Doppler shift $\Delta\nu_{\text{Doppler}}^{21} \approx \Delta\nu_{\text{Doppler}}^{12} \approx -2\nu_{\text{opt}} d\tau/dt$ is  

\begin{equation}
\frac{\Delta\nu_\text{{Doppler}}^{21} - \Delta\nu_{\text{Doppler}}^{12}}{\Delta\nu_{\text{Doppler}}^{12}} \approx \frac{\nu_{\text{AOM2}}-\nu_{\text{AOM1}}}{\nu_{\text{opt}}}\approx 5\times 10^{-8},
\end{equation}
where we have used $\nu_{\text{AOM1}}=37\,\text{MHz}$ and $\nu_{\text{AOM2}}=47\,\text{MHz}$.

We identify $\Delta\nu_{\text{Doppler}}^{21} \approx \Delta\nu_{\text{Doppler}}^{12}$ as the atmospheric noise, while $\Delta\nu_{\text{Doppler}}^{21} - \Delta\nu_{\text{Doppler}}^{12}$ contributes to the residual two-way noise.  

\section{Calculating the AM-PM effect}
\label{sec:AM-PM}
In our experiment, we record the voltage readout of the limiting amplifier, which is approximately proportional to the logarithm of the RF power, using an oscilloscope. From the recorded voltage, RF power and phase deviation are calculated. 

The conversion from voltage $V$ (in Volt) to RF power $P$ in dBm (used for Fig.~\ref{fig: am_pm}(b)) is linear for low RF power and accounts for a moderate compression at higher power
\begin{equation}
P(V) =
\begin{cases}
a_P^L + b_P^LV & V < V_1, \\
a_P^H + b_P^HV + c_p^HV^2 & V \geq V_1.
\end{cases}
\end{equation}

Here, the transition point $V_1$ corresponds to approximately $-32\,\text{dBm}$. The coefficients were determined by fitting data from a separate measurement, where the RF power after the photodetector and power splitter was measured using a spectrum analyzer.

For Fig.~\ref{fig: am_pm}(c), the conversion from voltage $V$ in Volt to phase deviation $\Phi$ in radians is given by a third-order polynomial at low RF power and an exponential at higher power
\begin{equation}
\Phi(V) =
\begin{cases}
a_{\Phi}^L + b_{\Phi}^LV + c_{\Phi}^LV^2 + d_{\Phi}^LV^3 & V < V_2, \\
a_{\Phi}^H + b_{\Phi}^H \exp(c_{\Phi}^HV) & V \geq V_2.
\end{cases}
\end{equation}

Here, the transition point $V_2$ corresponds to approximately $-38\,\text{dBm}$. The coefficients were determined by fitting the data plotted in Fig.~\ref{fig: am_pm}(a), where for the sake of simplicity only a projection to the $\Phi$-$P$ plane is shown, i.e. the polynomial and exponential functions were applied to the amplifier voltage readout and the corresponding RF power and phase, respectively, were then obtained for display purposes.

\begin{backmatter}

\bmsection{Acknowledgments}
We acknowledge the support by the Deutsche Forschungsgemeinschaft (DFG, German Research Foundation) under Germany's Excellence Strategy---EXC-2123 QuantumFrontiers---Project No. 390837967. The optical free-space link system was developed within the scope of the project QuNET, funded by the German Federal Ministry of Education and Research (BMBF) in the context of the federal government’s research framework in IT-security “Digital. Secure. Sovereign.”. We thank the “Stadtwerke Jena” for access to the free-space link infrastructure. Additionally, we extend our appreciation to Christopher Spiess for his invaluable administrative support.

\bmsection{Disclosures}
The authors declare no conflicts of interest.

\bmsection{Data Availability}
Data underlying the results presented in this paper are not publicly available at this time but may be obtained from the authors upon reasonable request.

\end{backmatter}

\bibliography{reference}

\end{document}